\documentstyle[12pt,graphicx]{article}

\newcommand{\Hamil}{{\cal H}}

\newcommand{\av}[1]{\langle{#1}\rangle}

\newcommand{\anticom}[2]{\{{#1},{#2}\}}

\newcommand{\ov}[2]{{\cal O}_{{#1}{#2}}}
\newcommand{\ovl}[2]{{\cal O}^{-1}_{{#1}{#2}}}
\newcommand{\Sov}[2]{{\cal S}_{{#1}{#2}}}
\newcommand{\genT}[1]{\tilde{T}_{#1}}
\newcommand{\deriv}[2]{\frac{{\rm d}#1}{{\rm d}#2}}

\def\bfk{{\bf k}}

\def\bfA{{\bf A}}

\def\dmu#1{{\rm d}#1}

\def\a#1{a_{#1\sigma}}
\def\adagger#1{a_{#1\sigma}^{\dagger}}
\def\c#1{c_{#1\sigma}}
\def\cdagger#1{c_{#1\sigma}^{\dagger}}
\def\l#1{l_{#1\sigma}}
\def\r#1{r_{#1\sigma}}
\def\ldagger#1{l_{#1\sigma}^{\dagger}}

\begin{document}
\title{\Large{On the non-orthogonality problem in the description of quantum devices}}
\author{\normalsize{\underline{Jonas Fransson}$^a$
\thanks{Corresponding author: Tel:+46 (0)18 4717308, Fax: +46 (0)18 511784, 
\underline{jonasf@fysik.uu.se} }, 
Olle Eriksson$^a$, B\"orje Johansson$^a$ and Igor Sandalov$^{a,b}$}\\
\normalsize{a-{\it Condensed Matter Theory group, Uppsala University}}\\
\normalsize{{\it Box 530, 751 21 Uppsala, Sweden}}\\
\normalsize{b-{\it Kirensky Institute of Physics, RAS, 660036 Krasnoyarsk, Russia.}} \\}
\date{}
\maketitle
\begin{abstract}
\baselineskip 24pt 
An approach which allows to include the corrections from
non-orthogonality of electron states in contacts and quantum dots is
developed.  Comparison of the energy levels and charge distributions
of electrons in 1D quantum dot (QD) in equilibrium, obtained within
orthogonal (OR) and non-orthogonal representations (NOR), with the
exact ones shows that the NOR provides a considerable improvement, for
levels below the top of barrier.  The approach is extended to
non-equilibrium states. A derivation of the tunneling current through
a single potential barrier is performed using equations of motion for
correlation functions. A formula for transient current derived by
means of the diagram technique for Hubbard operators is given for the
problem of QD with strongly correlated electrons interacting with
electrons in contacts. The non-orthogonality renormalizes the
tunneling matrix elements and spectral weights of Green functions
(GFs).
\end{abstract}
\vspace{5mm}
Keywords: \emph{non-orthogonality, current, tunneling}
\vspace{5mm}\\
Poster, your reference: MoP-06

\newpage
\baselineskip 24pt
{\bf Introduction:} In the description of tunneling processes through
quantum devices two approaches are useful: 1) the wave functions used
for calculation of the Green functions fulfil the boundary conditions
for the whole device, or 2) a subdivision of the system is made
\cite{pran63,meir92,inosh93,jauh94,oguri95,inosh97} and then the wave
functions corresponding to different subsystems are, in general, not
orthogonal to each other. The second approach is preferable if the
strength of the interactions in different regions of the system differ
considerably. In this framework tunneling arises due to the
non-orthogonality of the wave functions from different
subsystems. Prange \cite{pran63} found in his investigation of SNS-
and SIS-junctions, that an overcomplete non-orthogonal basis set,
allowing for the desirable separation leads to corrections from
overlap integrals of the same order as the tunneling
coefficients. Hence, it is essential to take the overlap between the
orbitals into account. We also note the conclusion of Svidzinskii
\cite{svid82}, that the tunneling Hamiltonian approach is
useful \emph{only} if one is interested in linear responses.

We present results of a different approach based on the diagram
technique for Hubbard operators within non-orthogonal basis
\cite{sand99}. Generalized to non-equilibrium states, the method still
allows for calculations in the language of model subsystems.

\begin{figure}[b]
\begin{center}
\includegraphics[width=10cm]{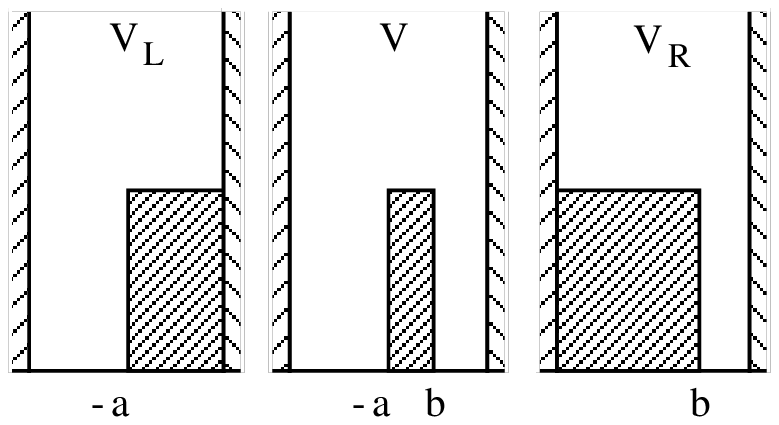}
\caption{Fransson et al.}
\label{boxes}
\end{center}
\end{figure}

{\bf Equilibrium:} Consider a finite box with hard walls containing a
barrier of finite height $V_0$ see Fig. \ref{boxes}; the Hamiltonian
is $\Hamil=\frac{p^2}{2m}+V$. We approximate this system by two
separate subsystems for which $\{\phi_p,\varepsilon_p\}$ and
$\{\phi_q,\varepsilon_q\}$ are complete ON eigensystems of the `left'
(L) and `right' (R) Hamiltonians $\Hamil_L=\frac{p^2}{2m}+V_L$, $p\in
L$ and $\Hamil_R=\frac{q^2}{2m}+V_R$, $q\in R$ corresponding to the
potentials $V_L=V\Theta(-x+b)+(V_0+V)\Theta(x-b)$ and
$V_R=(V_0+V)\Theta(-x-a)+V\Theta(x+a)$ see Fig \ref{boxes};
$\Theta(x)$ is the Heaviside step function.  Introduce the field
operators $\psi_L(x)=\sum_{p\sigma}\a{p}\phi_p(x)$ and
$\psi_R(x)=\sum_{q\sigma}\a{q}\phi_q(x)$; $\sigma$ denotes spin. The
exact field operator is expanded as $\psi=\psi_A+\psi_B$ where
$\psi_A=\psi_L+\psi_R$ and $\psi_B$ is a reminder. Assuming $\psi_B=0$
yields the approximate Hamiltonian
$\Hamil_A=\sum_{k\sigma}\varepsilon_k\adagger{k}\a{k}+\sum_{pq\sigma}\left(t_{pq}\adagger{p}\a{q}+H.c.\right)$
,where
$t_{kk'}=\int\phi_k^*\left(\frac{p^2}{2m}+V\right)\phi_{k'}\dmu{x}$;
$k=p,q$.  Then $t_{pq}=\ov{p}{q}\varepsilon_q+W_{pq}$, where
$\ov{p}{q}\equiv\int\phi_p^*\phi_q\dmu{x}$ defines the overlap matrix
and $W_{pq}\equiv\int\phi_p^*(V-V_R)\phi_q\dmu{x}$, and similarly for
the other matrix elements. Neglect the differences $W_{kk'}$ whenever
$k,k'$ belong to the same contact. The operators $\a{p}$ are defined
by
$\a{p}=\sum_{p'}\ovl{p}{p'}\int\phi_{p'}^*(x)\psi(x)\dmu{x}+\sum_{q'}\ovl{p}{q'}\int\phi_{q'}^*(x)\psi(x)\dmu{x}$,
and similarly for $\a{q}$. Then, the anti-commutation relations are
$\anticom{\a{k}}{\adagger{k'}}=\ovl{k}{k'}$, where $\ovl{k}{k'}$ is
element $kk'$ of the inversed overlap matrix.  Solutions of the Dyson
equation $g_A^{-1}=g_0^{-1}-\ovl{}{}W$ for $\ovl{}{}=I$ (OR), $I$ is
the identity operator, $\ovl{}{}\neq I$ (NOR) compared to the exact
solution are shown in Table \ref{tabell}. As seen, the improvement
achieved in NOR is considerable.  However, in the proximity of the
barrier height corrections from the reminder $\psi_B$ should be taken
into account when calculating the charge distribution, even though the
energy levels estimated by NOR still are much better.

{\bf Transient current:} Transient current can be calculated
\cite{meir92} from $-e\frac{d}{dt}\av{N_L}$, where
$N_L$ is the number of carriers in the cylinder with top and bottom
areas $S$ and axis parallel to current. The tunneling matrix elements
now contain the vector potential, so, in the Hamiltonian $\Hamil_A$
substitute $t_{kk'}$ by $T_{kk'} =\langle k|(\bfk -(e/c)\bfA(t))^2 +
V|k'\rangle$, the non-equilibrium tunneling coefficients.  Whenever
$k,k'$ belong to the same contact we approximate $T_{kk'}$ by its
corresponding equilibrium value and neglect the differences $W_{kk'}$,
yielding for example $T_{pp'}\rightarrow\varepsilon_p$.

Via the transformation $\a{p(q)}=e^{-i\varphi_{L(R)}(t)}\l{p}(\r{q})$,	
which allows to introduce current states, we derive a tunneling
current $J\sim |T|^2$ as a function of the applied voltage
$eV=\mu_L-\mu_R\approx\frac{\varphi(t)-\varphi(t')}{t-t'}$, where the
phase $\varphi=\varphi_L-\varphi_R$. Putting
$\ovl{k}{k'}=\delta_{kk'}$ when $k,k'$ belong to the same contact, is
consistent with the approximation. Note that with this assumption
$\av{N_L}=\sum_{p\sigma}\Sov{p}{}\av{n_{p\sigma}}$, where
$n_{p\sigma}=\adagger{p}\a{p}$ and $\Sov{p}{}=\Sov{p}{p}$ is a part of
the overlap matrix, obtained by integration over the volume in which
$-e\frac{d}{dt}\langle \hat{N}_L \rangle $ is calculated. The
equations of motion for $\av{n_{p\sigma}}$ is
$\deriv{}{t}\av{n_{p\sigma}}=2{\rm
Im}\sum_{q}\genT{pq}e^{i\varphi(t)}\av{\ldagger{p}\r{q}}$ where
$\av{\ldagger{p}\r{q}}=\genT{pq}(f_p-f_q)\frac{(-i)^2}{\Delta_{qp}+eV-i\delta}$,
$\Delta_{qp}=\varepsilon_q-\varepsilon_p$,
$\genT{pq}=T_{pq}+\ovl{p}{q}\varepsilon_q$ in the given approximation
and $f$ is the Fermi function.  The resulting current formula in terms
of densities of states $N_L(\mu)$ and $N_R(\mu)$ then is 
$J=V\cdot4\pi e^2S\Sov{}{}|\genT{}|^2N_L(\mu)N_R(\mu)\equiv VR^{-1},$ 
(the factor 2 is due to spin). Thus, in this approximation the only change required
is replacing $T\rightarrow \tilde{T}$.

{\bf Strong correlations in quantum dot:} Physically, the coupling
arises due to overlap of wave functions. When the system is close to
the regime of Coulomb blockade, the overlap is small and the
interaction between subsystems is much weaker than the one inside the
QD, {\em i.e.} the matrix element of tunneling $|T|<<\Delta
_{\bar{a}}=E_{\Gamma _{n+1}}-E_{\Gamma _n}$; here $E_{\Gamma _n}$ is
an eigenvalue of the Hamiltonian of the QD, $H_d|\Gamma
\rangle =E_\Gamma |\Gamma \rangle $; $a=[\Gamma _n,\Gamma _{n+1}]$ 
is a fermion-like transition ($\bar{a}=[\Gamma _{n+1},\Gamma _n]$)
which is described by Hubbard operator $X^{\Gamma _n\Gamma
_{n+1}}\equiv X^a$. In this situation the states of the QD will be
perturbed only slightly, therefore QD should be described by
many-electron states. Here we will demonstrate how the technique
developed in ref. \cite{sand99} for correlated electron systems for
thermodynamics can be extended to non-equilibrium phenomena. Since the
Hamiltonian includes {\em strong} Coulomb interactions inside QD, the
terms of the kind $v_{k\sigma ,\mu _2\mu _3\mu _4}l_{k\sigma
}^{\dagger }d_{\mu _2}^{\dagger }d_{\mu _3}d_{\mu _4}=v_{k\sigma ,\mu
_2\mu _3\mu _4}(d_{\mu _2}^{\dagger }d_{\mu _3}d_{\mu _4})^al_{k\sigma
}^{\dagger }X^a\equiv T_{k\sigma ,a}^{Coul}l_{k\sigma }^{\dagger
}X^a$, also contribute to the process of electron tunneling from the
left contact to the QD and should not be decoupled in Hartree-Fock
fashion. We include these interactions to the Hamiltonian of coupling
(see ref. \cite{sand99}). The total Hamiltonian then is:
\begin{eqnarray*}
\Hamil =\sum_{k\sigma} \varepsilon _{k\sigma }\cdagger{k}\c{k}+\sum_{\Gamma}
E_\Gamma X^{\Gamma \Gamma }+\sum_{k\sigma,a}(T_{k\sigma ,a}\cdagger{k}X^a+H.c.)
\end{eqnarray*}
where $k=p,q$ ($p\in L$ and $q\in R$). Any $X$-operator can be
rewritten in terms of products of single-electron operators
$d,d^{\dagger }$ of QD and, therefore, using $\{\c{k},d_\mu ^{\dagger
}\}={\cal O}_{k\sigma }^{-1}{}_{,\mu }$ one can show \cite {sand99}
that $\{\c{k},X^{\bar{a}}\}={\cal O}_{k\sigma }^{-1}{}_{,\mu }(d_\mu
)^b\varepsilon _\xi ^{b\bar{a}}X^\xi $, where $\xi $ is a Bose-like
transition, and $\varepsilon _\xi ^{b\bar{a}}$ defines commutation
relations between $X$-operators in the uncoupled system,
$\{X^b,X^{\bar{a}}\}=\varepsilon _\xi ^{b\bar{a}}X^\xi .$ Strictly
speaking, when the coupling is switched on, the many-electron states
also become non-orthogonal to each other but the corrections contain
higher order products of $\ovl{k\sigma,}{\mu }$. Hence, we neglect
these corrections since we are only interested in first order with
respect to transparancy. Following ref. \cite{meir92} we calculate the
contribution to the current from `left' electrons
$J_{tr}^{(l)}=-2eS{\cal S}{\rm
Re}\sum_{k\sigma,a}\tilde{T}_{k\sigma,a}^{(l)*}G_{k\sigma,\bar{a}}^{<}(tt)$,
where $\tilde{T}_{k\sigma
,a}^{(l)}=T_{k\sigma,a}+\ovl{k\sigma,}{\mu}(d_\mu )^b\varepsilon
_a^{b\Gamma }E_\Gamma $.  The additional term comes from the
anti-commutation of $\c{k}$ and the QD Hamiltonian
$\Hamil_D$. Expressing the current in terms of retarded and `lesser'
GFs of the QD, $G_{a\bar{a}}^R$ and $ G_{a\bar{a}}^{<}$ respectively,
yields:
\[
J_{tr}^{(l)}=2eS{\cal S}{\rm Im}\sum_{k\sigma,a}\left\{
\genT{k\sigma,a}^{(l)*}\ovl{k\sigma}{,\mu}(d_{\mu})^aP^af_L(\varepsilon _{k\sigma })
-|\tilde{T}_{k\sigma ,a}^{(l)}|^2
[G_{a,\bar{a}}^{<}(\varepsilon _{k\sigma })
+f_L(\varepsilon _{k\sigma })
G_{a,\bar{a}}^R(\omega )_{|\varepsilon _{k\sigma }}]\right\}
\]
For a transition $a=[\gamma,\Gamma]$ the expectation value
$P^a=\av{\anticom{X^{\gamma\Gamma}}{X^{\Gamma\gamma}}} =N^\gamma
+N^\Gamma$, \emph{i.e.} it is a sum of the population numbers
corresponding to the transition. They should be found from $\int
G_{a,\bar{a}}^{<(>)}(\omega ) d\omega$. The system for
$G_{a,\bar{a}}^{<(>)}(\omega )$ is very cumbersome and therefore not
given here. However, physics is seen from the form of the
retarded GF: $G_{a,\bar{a}}^{R}(\omega )=\frac{[1+\gamma
(\omega )]P^a}{\omega + i\delta -\Delta _{\bar{a}}-\Gamma _a(\omega
)P^a},$ where $\Gamma _a(\omega )=\Gamma _a^{(l)}(\omega )+\Gamma
_a^{(r)}(\omega ) $, $\gamma _a(\omega )=\gamma _a^{(l)}(\omega
)+\gamma _a^{(r)}(\omega )$, and the width $\Gamma _a^{(l)}(\omega
)=\sum \tilde{T}_{p\sigma ,a}^{(l)*}g_{p\sigma }(\omega
)\tilde{T}_{p\sigma ,a}^{(l)}$. $g$ is bare GF of `left' electrons,
$\gamma _a^{(l)}(\omega )=\tilde{T}_{p\sigma ,a}^{(l)*}g_{p\sigma
}(\omega )\ovl{p\sigma,}{\mu }(d_\mu )^bP^b$ and in $\Gamma
_a^{(r)}(\omega )$, $\gamma _a^{(r)}(\omega )$ summation is over $
q,\sigma $. Thus, each single-electron intra-dot transition acquires
width, which depends on the overlap of the wave function of conduction
electron in the left(right) contact with energy near Fermi level $\mu
_L(\mu _R)$ with those in-dot orbitals in transition $a$.

In conclusion we have shown that, the separation of a device into
auxiliary subsystems unavoidably leads to eigenbases non-ortogonal to
one another.  This results in additional contributions to matrix elements
of tunneling and in redistribution of spectral weights since part of
charge is in `intermediate' state. Precision of calculations is improved
even in the most `dangerous' region at the top of the barrier.

\newpage
\begin{table*}[h]
\begin{center}
\begin{tabular}{ccc}
{\bf Exact}&OR&{\bf NOR} \\ 
\hline{\bf 0.28} & 0.012 & {\bf 0.29}\\ 
\hline{\bf 1.06} & 0.80 & {\bf 1.08}\\ 
\hline{\bf 2.28} & 2.07 & {\bf 2.30}\\ 
\hline{\bf 4.04} & 3.31 & {\bf 4.18} 
\end{tabular} 
\end{center}
\caption{Energy levels (mHartree) for barrier width 1 Bohr and height 4.1 mHartree.}
\label{tabell} 
\end{table*} 

\newpage

\begin{center}
\Large{\bf Caption}
\end{center}
\begin{itemize}
\item[\bf{Fig. 1}] The assumed potentials - `left', `exact' and `right'.
\end{itemize}

\newpage

\end{document}